\documentclass[11pt, reqno]{amsart}

\usepackage{amsmath,amsfonts}
\usepackage{latexsym,amssymb}
\usepackage{amsthm}
\usepackage{bm}
\usepackage[svgnames,dvipsnames]{xcolor}
\usepackage{geometry}

\newcommand{\dd}{\mathtt{d}}
\newcommand{\grad}{\bm{\nabla}}

\begin{document}

\title[]{Comment on work of Yang ang Nevels ``Direct, analytic solution for the electromagnetic vector potential in any gauge'' (arXiv:2507.02104)}

\author[]{Vladimir Onoochin}

\begin{abstract}
In the Comment the procedure for obtaining a solution for the electromagnetic potential, presented in the cited work arxiv.org:2507.02104, is analyzed. It is shown that the solution obtained by the authors is based on some mathematically illegal operations. This argument is supported by a counterexample to Eq. (10) of the cited work. Direct calculations of the potentials in the Lorenz and Coulomb gauges show that Eq. (10) is not satisfied.
\end{abstract}

\maketitle

In a recent work by Yang and Nevels~\cite{YN} the authors present a derivation of an expression for the vector potential for an arbitrary time-dependent charge-current distribution. Also the authors make significant statement that they are able to obtain this expression without adding the gauge condition. If their procedure for deriving the expression for the vector potential is correct, this represents a simple and elegant proof that the potentials $\Phi$ and ${\bf A}$, calculated in any gauge, give true expressions for the electromagnetic fields. 

One can see that in deriving the final expression for ${\bf A}_2$, the part of the vector potential created by non-local source, the authors do not use any physical arguments but only mathematical transformations. Unfortunately, these transformation are made without the necessary mathematical rigor which makes their derivation procedure to be questionable. Let us demonstrate this.

The final expression for ${\bf A}$, Eq.~(10) of~\cite{YN}, contains unknown function, $\Phi({\bf r},t)$. This function cannot be found from the initial equations, Eqs.~(1)-(3) of~\cite{YN},
\begin{eqnarray}
	\bm{\nabla}^2 \Phi ({\bf r}, t) + \frac{1}{c}\frac{\partial}{\partial t} [ \bm{\nabla}\cdot {\bf A}({\bf r}, t) ] 
	= - 4 \pi \rho ({\bf r}, t) \,, 	\label{eq-A2}\\
	\left( \bm{\nabla}^2 -  \frac{1}{c^2} \frac{\partial^2}{\partial t^2}  \right) {\bf A}({\bf r}, t)  = -  \frac{4\pi}{c}  {\bf J}({\bf r}, t)
+	\bm{\nabla}\left( \bm{\nabla} \cdot {\bf A}({\bf r}, t) +  \frac{1}{c}  \frac{\partial \Phi ({\bf r}, t) }{\partial t} \right) \,,
	\label{eq-A3}
\end{eqnarray}
 Inserting the solution for the vector potential into Eq.~\eqref{eq-A2} one obtains,
\begin{equation}
	\bm{\nabla}^2 \Phi ({\bf r}, t) +\frac {1}{ c}\frac {\partial }{\partial t} [ \grad \cdot {\bf A}_1({\bf r}, t) ]  + \frac{\partial }{ \partial t} \grad\int
	\left[\Phi_c({\bf r},t) - \Phi({\bf r},t) \right]\dd t 	= - 4 \pi \rho ({\bf r}, t),
	\label{eq-phi}
\end{equation}
where $\Phi_c({\bf r},t)$ is the scalar potential in the Lorenz gauge. One can see that the term $\bm{\nabla}^2 \Phi ({\bf r}, t) $ is compensated by the term  in the time integral. So one obtain
\[
\bm{\nabla}^2 \Phi _c({\bf r}, t) + \frac{1}{c}\frac {\partial }{\partial t} [ \grad \cdot {\bf A}_1({\bf r}, t) ] =- 4 \pi \rho ({\bf r}, t)\,,
\]
that can be reduced to the wave equation for $\Phi_c$ by using the Lorenz gauge condition. Although the solution for the vector potential satisfies Eq.~\eqref{eq-A2}, it does not allow us to find the scalar potential $\Phi$, since the terms containing this quantity are removed from the differential equation.

Based on this property of Eq.~\eqref{eq-phi}, the authors conclude that the expression for the vector potential, Eq.~(10) of~\cite{YN}, is correct for the scalar potential $\Phi$ calculated in any gauge (Secs. 3--5 of the cited work).

If Eq.~(10) of~\cite{YN} is correct, this equation can be used to demonstrate that the potentials calculated in {\it any gauge} yield {\it identical} expressions for the electromagnetic fields. In the other words,  all gauges are equivalent. For such a demonstration it is sufficient to calculate the partial time derivatives of both sides in Eq.~(10)~of~\cite{YN},
\begin{equation*}
\begin{split}
\frac{1}{c}\frac{\partial {\bf A}}{\partial t}= \frac{1}{c}\frac{\partial {\bf A}_1}{\partial t} +\grad\Phi_c - \grad \Phi \quad \to \\
\to \quad \grad\Phi_c +\frac{1}{c}\frac{\partial {\bf A}_1}{\partial t} = \grad\Phi+ \frac{1}{c}\frac{\partial {\bf A}}{\partial t}
=-{\bf E}
\end{split}
\end{equation*}
However, the procedure of deriving the final expression for the vector potential is based on some incorrect mathematical operations. It makes questionable the final expression itself. Let us demonstrate this by presenting the counterexample to Eq.~(10) of~\cite{YN}. This counterexample will be obtained in the following way:\newline
$\square$ First it will be calculated ${\bf A}_2$ from the expression
\begin{equation}
{\bf A}_{2}^\prime ({\bf r}, t) = c \bm{\nabla} \int \left[ \Phi_{L} ({\bf r}, t) - \Phi_{Cl} ({\bf r}, t) \right] dt, 
\label{eq-A8}
\end{equation}
where $\Phi_L$ and $\Phi_{Cl}$ are the scalar potentials in the Lorenz and Coulomb gauges.\newline
$\square$ Second, ${\bf A}_2$, a part of the vector potential in the Coulomb gauge created by non-local source, will be calculated following the standard approach, {\it i.e.} solving the wave equation for the vector potential, Eq.~(6.24) of~\cite{JDJ}. \newline
$\square$ Then the obtained result will be compared with ${\bf A}_2^\prime$ which is obtained from Eq.~\eqref{eq-A8}.

If one intends to calculate ${\bf A}_2^\prime$ in the closed form, {\it i.e.} the expression that does not contain integrals which should be computed, the scalar potential $\Phi_{L}$ ($\Phi_c$ in Yang-Nevels' notation) must be expressed in the present time variables. It is quite impossible to compute the time integral of the function depending on the retarded time. Therefore, one needs to use the scalar potential created by a classical charge moving with constant velocity or acceleration. For verification of the formula~\eqref{eq-A8}, let us calculate the scalar potentials created by a charge  set suddenly from rest at ${\bf R}=0$ for $t<0$ into uniform motion in the $x$ axis with the velocity $v=const$. If the scalar potentials are \lq{}detected\rq{} in the same axis, $\Phi_{L}(X,t) = \Phi_{Cl}(X,t)$ and ${\bf A}_2^\prime(X,t)=0$.

Solution of Eq.~(6-24) of~\cite{JDJ} for ${\bf A}_2$, re-written in the Gaussian units, is
\begin{equation}
{\bf A}_2({\bf R},t)=\frac{1}{c}\int \frac{\left[\partial_t\grad\Phi_{Cl}({\bf r},t') \right]}
{ \vert{\bf R}-{\bf r}\vert}\dd {\bf r}\,.
\end{equation}
where the term in the square brackets is expressed in the retarded time $t'=t-\vert{\bf R}-{\bf r}\vert/c$. Also it should be noted that in calculation of the derivatives of $\Phi_{Cl}({\bf r},t)$, these derivatives are calculated for the time variable $t$ and after it, this variable is changed to $t'$.

For the potentials \lq{}detected\rq{} in the $x$ axis, the above integral can be written  as
\begin{equation}
{A}_{2,x}({\bf R},t)=\frac{1}{c}\int
\left[\dfrac{\partial}{\partial t}\dfrac{\partial}{\partial x}\dfrac{q}{\sqrt{(x -vt' )^2+y^2+z^2}}\right]\frac{\dd {\bf r}}
{ \sqrt{(X-x)^2+y'^2+z'^2} }\,,
\end{equation}
with $t'=t- \sqrt{(X-x )^2+y^2+z^2}/c$. This integral cannot be computed in the closed form. But if one accepts $X=0$ (the potentials are  \lq{}detected\rq{} in the origin of the coordinates), $t'=t- \sqrt{x ^2+y^2+z^2}/c=t-r/c$, and the above integral is transformed to 
\begin{equation}
A_{2,x}(0,t)=\frac{1}{c}\int
\left[\frac{\partial}{\partial t}\frac{\partial}{\partial x}\dfrac{q}{\sqrt{(x -vt' +vr/c)^2+y^2+z^2}}\right]\frac{\dd {\bf r}}{\sqrt{x^2+y^2+z^2}}\, .\label{eq-Asiml}
\end{equation}
It should be noted that calculation of partial derivatives of the term in the square brackets gives singularity so these derivatives are computed using the Frahm's identity~\cite{Frahm}
\begin{equation*}
\begin{split}
&\frac{\partial}{\partial t}\frac{\partial}{\partial x}\dfrac{q}{\sqrt{(x -vt +vr/c)^2+y^2+z^2}}=\\
&=\frac{4\pi qv}{3}\delta(x -vt +vr/c)\delta(y)\delta(z)-
\frac{qv [2(x-vt+vr/c)^2-y^2-z^2]}{(x-vt+vr/c)^2 + y^2+z^2)^{5/2}}\,.\label{Frahm}
\end{split}
\end{equation*}
The integral with the delta-functions is easily computed, which gives
\begin{equation}
I_1=\frac{4\pi qv}{3c}\int \frac{\delta(x -vt +vr/c)\delta(y)\delta(z) }{\sqrt{x^2+y^2+z^2}}\dd x\dd y\dd z = \frac{4\pi q}{3ct}\,.
\end{equation}
The integral with the second term in the {\it rhs} of Frahm's identity can be calculated in the spherical coordinates $r,\,\theta,\,\phi$
\begin{equation}
\begin{split}
I_2=-\frac{qv}{c}\int \frac{2(x-vt+vr/c)^2-y^2-z^2}{(x-vt+vr/c)^2 + y^2+z^2)^{5/2}}\frac{\dd x\dd y\dd z}{\sqrt{x^2+y^2+z^2}}= \\
= -\frac{2\pi qv}{c}\int\limits_0^{\infty}r\dd r\int\limits_0^{\pi}\frac{2(r\cos\theta-vt+vr/c)^2-r^2\sin^2\theta}{(r\cos\theta-vt+vr/c)^2+r^2\sin^2\theta)^{5/2}}
\sin\theta\dd\theta\,.
\end{split}
\end{equation}
The integral over the angular variable $\theta$ can be computed by means of {\it Mathematica} software,
\[
\int\limits_0^{\pi}\frac{2(r\cos\theta-vt+vr/c)^2-r^2\sin^2\theta}{(r\cos\theta-vt+vr/c)^2+r^2\sin^2\theta)^{5/2}}
\sin\theta\dd\theta = \frac{2(1-{\rm sgn}(r-vt+vr/c))}{v^3t^3}\,,
\]
where ${\rm sgn}(\xi)$ is a signum function of the real argument $\xi$. Thus, calculation of this integral with respect to $r$ gives
\begin{equation}
I_2=-\frac{4\pi q}{v^2ct^3}\int \limits_0^{\infty} r\dd r \left[1-{\rm sgn}(r-vt+vr/c) \right]=-\frac{8\pi q}{vct^3}\int \limits_0^{\frac{vct}{c+v}} r\dd r =-
\frac{4\pi cq}{(c+v)^2t}\,.
\end{equation}
Therefore,
\begin{equation}
A_{2,x}(0,t)=I_1+I_2=-\frac{4\pi q}{ct}\left[ \frac{c^2}{(c+v)^2}- \frac{1}{3}\right]\neq 0\,.
\end{equation}
Thus, the value of the vector potential ${\bf A}_2$ is not equal to the value of the vector potential which can be obtained from Eq.~(10) of~\cite{YN}. It means that the derivation of the vector potential presented in the cited work is questionable.\\

Now it would be desirable to give some explanations why the final expression, Eq.~(10) of~\cite{YN} is incorrect. It is known that the inhomogeneous wave equation, as other partial differential equations with the {\it rhs} is solved by means of the Duhamel principle, when the inhomogeneous problem is considered as a set of homogeneous problems each starting afresh at a different time slice $t = t_0$. By linearity, one can integrate the resulting solutions through time $t_0$ and obtain the solution for the inhomogeneous problem. In this approach, the {\it rhs} of the equation is considered as a set of initial conditions. It means that the unknown variable of the equation and the {\it rhs} of this equation are different mathematical quantities. But one can see that the authors use the vector potential in Eq.~(3) of~\cite{YN} in both sides of the equation, as unknown variable and as the source. It is mathematically illegal operation.

Although then the authors change ${\bf A}$ in the {\it rhs} of Eq.~(3) to the scalar potential and the charge density, it cannot make the illegal operation to be legal. In this change the authors use the equation for ${\bf A}$. But the equation is not the equality.

It should be noted that in the standard approaches to solve the system of the Maxwell equation, Ch.~6.2 of~\cite{JDJ}, the vector potential is not used as a source. Although the scalar potential is used as a source in the equation for the vector potential, but this operation is legal since the scalar potential has been determined at the previous step of solving.

Finally, it should be concluded that the authors have not obtained the direct, analytic solution for the electromagnetic vector potential in any gauge in the mathematically rigorous way. This is confirmed by presenting the counterexample to their Eq.~(10).

\end{document}